\documentclass[twocolumn,aps,superscriptaddress,showpacs,floatfix]{revtex4}
\usepackage{amssymb}
\usepackage{amsmath}
\usepackage{graphicx}
\usepackage[normalem]{ulem}
\usepackage[dvips]{color}
\usepackage{bm}
\usepackage{longtable}
\usepackage{array}
\usepackage{etoolbox}
\usepackage{setspace}
\usepackage[colorlinks,linkcolor=blue,anchorcolor=blue,citecolor=blue,urlcolor=blue]{hyperref}

\renewcommand\sout{\bgroup \color{red} \ULdepth=-.5ex \ULset}

\begin{document}

\title{Systematic study of $\alpha$ decay half-lives for even-even nuclei within a two-potential approach}

\author{Xiao-Dong Sun}
\affiliation{School of Math and Physics, University of South China, 421001 Hengyang, People's Republic of China}
\author{Ping Guo}
\affiliation{School of Math and Physics, University of South China, 421001 Hengyang, People's Republic of China}
\author{Xiao-Hua Li\footnote{%
Corresponding author:lixiaohuaphysics@126.com }}
\affiliation{School of Nuclear Science and Technology, University of South China, 421001 Hengyang, People's Republic of China}
\affiliation{Cooperative Innovation Center for Nuclear Fuel Cycle Technology $\&$ Equipment, University of South China, 421001 Hengyang, People's Republic of China}


\begin{abstract}

$\alpha$ decay is a common and important process for natural radioactivity of heavy and superheavy nuclei. The $\alpha$ decay half-lives for even-even nuclei from Z=62 to Z=118 are systematically researched based on the two-potential approach with a quasi-stationary state approximation. To describe the deviations between experimental half-lives and calculated results due to the nuclear shell structure, a hindrance factor related with $\alpha$ particle preformation probability is introduced. Our results can well reproduce the experimental data equally to the density-dependent cluster model and the generalized liquid drop model. We also study the isospin effect of nuclear potential in this work. Considering the isospin effect the calculated results improved about 7.3$\%$.

\end{abstract}

\maketitle

\section{Introduction}

Transuranic elements follow the trend that their half-lives decrease as atomic numbers increase until the next nuclear shell appears. Synthesized atoms of the most recently discovered 117 element have lasted some tens of microseconds~\cite{Oga12,Khu14}, gradually approaching the island of stable superheavy element~\cite{Hof00}. One of useful ways to confirm the superheavy elements is to discriminate specific $\alpha$ particles emanated from itself as well as its $\alpha$ decay chain nuclei. Landing the island of stability and other interesting discoveries, such as the triplet shape coexistence~\cite{And00} and extremely long $\alpha$ decay half-life nuclide $^{209}Bi$~\cite{De03} and so on, make experimental and theoretical researches on $\alpha$ decay becoming one of hot topics again.

In 1928, Gamow and Condon~\cite{Gam28} and Guerney~\cite{Gur28} had independently put forward the quantum tunnel theory, which successfully estimates the probability of an $\alpha$ particle tunnelling through the Coulomb barrier. The process of barrier tunnelling (penetration) is one important assumption for $\alpha$ decay. The other is that $\alpha$ particle cluster is prone to forming on the surface of the parent nucleus. The existing problem is that before the $\alpha$ particle in bound state collided with the barrier, we know little of the $\alpha$ particle how to form and motion inside the parent nucleus. The difficulties come from the complicated structure of the quantum many body system, e.g. the collective deformation, the fundamental excitations and the nuclear shell closure, and the uncertain of nuclear potential between the $\alpha$ particle and remaining nucleus.

The traditional methods~\cite{Poe79,Gur87,Buc90,Gon93,Roy00,Xu06,Zha06,Moh06,Den15,San15,Poe12, Zde13,Tav05,Guo15,Del10}such as the WKB approximation, and empirical formulas~\cite{Gei11,Ren12,Del09,Wan15} are constantly evolving. Among these methods the microscopic double-folding model adopting density dependent M3Y force and the liquid drop model adopting the proximity potential have been researched frequently. Other methods are also developed for $\alpha$ decay, e.g. the coupled-channel method is used to interpret the fine structure of $\alpha$ decay~\cite{Ni10}. The first empirical formula for $\alpha$ decay, Geiger-Nuttall(GN) law in 1911~\cite{Gei11}, relates the $\alpha$ decay half-lives with the decay energy $Q$, and its microscopic interpretation will improves the accuracy of GN law~\cite{Qi14}. These calculations are very successful for $\alpha$ decay. In general, the absolute $\alpha$ decay constant is determined by the preformation probability, the assault frequency and the penetration probability. It is arduous to obtain the actual wave function of parent nucleus and decay state, thus the preformation probability is ambiguous. The shell effect controls the trend that the preformation probability abruptly decreases in the vicinity of the nucleon magic number. Fortunately on the one hand the effective preformation factor can be extracted from the ratios of the experimental $\alpha$ decay half-lives to the calculated penetration probability~\cite{Zha09,Qi09,Gan09,Qia13,Guo15,Sei15}. On the other hand a microscopic shell model plus cluster component can provides the preformation probability successfully~\cite{Var92,Lov98,Del13,Bet12,War15}.

In this article we focus on predicting $\alpha$ decay half-life more accurately and studying isospin effect of the nuclear potential. We adopt the two potential approach with a quasi-stationary state approximation~\cite{Gur87}, and draw on the analytic expression for $\alpha$ particle preformation probability in Ref.~\cite{Guo15} to estimate variation of the preformation probabilities with the number of valence nucleon. The model parameters are obtained by fitting 164 $\alpha$ decay half-lives of even-even nuclei taken from the newest nuclear property table NUBASE2012~\cite{Aud12NUBASE}. Based on the above model we systematically calculate the half-lives of even-even nuclei, thereafter we study the relations of depth and diffuseness of the nuclear potential to isospin by fitting the experimental data. 

This article is organized as follows. In Sec.II the theoretical framework of the calculation of $\alpha$ decay half-lives and analytic expression for $\alpha$ particle preformation probability are briefly described. In Sec.III we present numerical results, discussion for the hindrance factor and the isospin effect of nuclear potential. A brief summary is given in Sec.IV.

\section{Theoretical framework}

The half-life $T_{1/2}$ for $\alpha$ decay could be determined by $\alpha$ decay width $\Gamma$ or decay constant $\lambda$. It can be written as
\begin{eqnarray}\label{1}
T_{1/2}=\frac{\hbar ln2}{\Gamma}=\frac{ln2}{\lambda}.
\end{eqnarray}
The decay constant $\lambda$ depending on the $\alpha$ particle preformation probability $P_{\alpha}$, the penetration probability $P$ and the normalized factor $F$, which represents the collision probability or assault frequency, can be expressed as
\begin{eqnarray}\label{2}
\lambda=\frac{P_{\alpha}FP}{h},
\end{eqnarray}
where $h=\frac{T^{exp}_{1/2}}{T^{cal}_{1/2}}$ is defined as hindrance factor. The superscript $exp$ and $cal$ represent experimental data and calculated values, respectively. The normalized factor $F$, which is given by the integration over the internal region~\cite{Gur87}, can be written as
\begin{eqnarray}\label{3}
F\int_{r_1}^{r_2}\frac{\mathit{d}r}{2k(r)}=1,
\end{eqnarray}
where $r$ is the mass center distance between the preformed $\alpha$ particle and the daughter nucleus. The $r_1$, $r_2$ and following $r_3$ are the classical turning points. $k(r)=\sqrt{\frac{2\mu}{\hbar^2}\mid Q_\alpha-V(r)\mid}$ is the wave number. $\mu$ is the reduced mass of the $\alpha$ particle and daughter nucleus in the center of mass coordinate. $V(r)$ and $Q$ are the height of $\alpha$-core potential and $\alpha$ decay energy, respectively. The penetration probability $P$, which is calculated by WKB approximation, can be expressed as
\begin{eqnarray}\label{4}
P=exp[-\frac{2}{\hbar}\int_{r_2}^{r_3}k(r)\mathit{d}r].
\end{eqnarray}
The classical turning points satisfy the condition $V(r_1)=V(r_2)=V(r_3)=Q$. In the inner region $(r_1<r<r_2)$ the strong interaction commands the state of the preformed $\alpha$ particle, while in the outer region $(r_2<r<r_3)$ the electromagnetic interaction plays a major role.

The potential between the preformed $\alpha$ particle and the daughter nucleus, including nuclear, Coulomb and centrifugal potential barrier, can be written as
\begin{eqnarray}\label{5}
V(r)=V_N(r)+V_C(r)+V_l(r).
\end{eqnarray}
Where $V_N(r)$ represents nuclear potential, which is critical and uncertain for $\alpha$ decay. In this work, we choose a type of $cosh$ parameterized form for nuclear potential~\cite{Buc92}. It can be expressed as
\begin{eqnarray}\label{6}
V_N(r)=-V_0\frac{1+cosh(R/a)}{cosh(r/a)+cosh(R/a)},
\end{eqnarray}
where $V_0$ and $a$ are parameters of the depth and diffuseness for the nuclear potential, respectively. $V_C(r)$ is the Coulomb potential and is taken as the potential of a uniformly charged sphere with sharp radius $R$, which can be expressed as
\begin{eqnarray}\label{7}
V_C(r)= \left \{
\begin{aligned}
\frac{Z_dZ_{\alpha}\mathit{e}^2}{2R}[3-(\frac{r}{R})^2]~~~r<R\\
\frac{Z_dZ_{\alpha}\mathit{e}^2}{2r}~~~~~~~~~~~~~~~~r>R,
\end{aligned}
\right.
\end{eqnarray}
where $Z_d$ and $Z_{\alpha}$ are proton number of the daughter nucleus and the $\alpha$ particle, respectively. The sharp radius of interaction $R$ is given by
\begin{eqnarray}\label{8}
R=1.28A^{1/3}-0.76+0.8A^{-1/3}.
\end{eqnarray}
This empirical formula is commonly used to calculate $\alpha$ decay half-lives~\cite{Roy00}, which derived from the nuclear droplet model and the proximity energy. $V_l(r)=\frac{l(l+1) \hbar^2}{2\mu r^2}$ is centrifugal potential, where $l$ is the orbital angular momentum taken away by $\alpha$ particle. In general, only the favored transitions ($l=0$) take place for $\alpha$ decay of even-even nuclei~\cite{Guo15}, and then $V_l(r)=0$. 

The hindrance factor $h$ reflects the deviations between the calculated half-lives $T_{1/2}^{cal}$ with constant preformation probability $P_\alpha$ and experimental half-lives $T_{1/2}^{exp}$. It will systematically varies due to the nuclear shell effect. The trend of hindrance factor $h$ can be estimated by the simple formula with five parameters proposed by Zhang \textit{et al} to research the preformation probaiblity of $\alpha$ particle varies in the different nuclear shells~\cite{Zha09,Guo15}. The hindrance factor can be given by
\begin{gather}\label{9}
log_{10}h=a+b(Z-Z_1)(Z_2-Z)+c(N-N_1)(N_2-N)\nonumber\\
+dA+e(Z-Z_1)(N-N_1),
\end{gather}
where $Z$, $N$ and $A$ are the proton, neutron and mass numbers of parent nucleus. $Z_1$ and $Z_2$ ($N_1$ and $N_2$) are the proton (neutron) magic numbers around $Z$ ($N$). $a$, $b$, $c$, $d$ and $e$ are the adjustable parameters. In eq.(9), the first and fourth terms describe the magnitude and the trend of the preformation probability with the increasing mass number, the second and third terms show a parabolic dependence of $log_{10}h$ as a function of the valence proton (neutron) number, the last term relates to the integrated valence neutron-proton interaction strength~\cite{Guo15,Gan09}.

\section{Results and discussions}

\subsection{Systematic calculation of half-lives}

We calculate the half-lives for even-even nuclei $\alpha$ transition between ground states of parent nuclei and daughter nuclei within the two-potential approach. The experimental data of $\alpha$ decay energy $Q_\alpha$ and half-lives $T^{exp}_{1/2}$ are taken from AME2012~\cite{Wan12,Aud12AME} and NUBASE2012~\cite{Aud12NUBASE}, respectively. The adjustable parameters, depth $V_0$ and diffuseness $a$ of nuclear potential and average value of preformation probability $P_\alpha$, are fitted by minimizing the total square deviation $\Delta$, which is defined as
\begin{eqnarray}\label{10}
\Delta=\sum_{i=1}^{N}(log_{10}T_{1/2}^{cal}-log_{10}T_{1/2}^{exp})^2.
\end{eqnarray}
All the experimental data of 164 even-even nuclei from Z=62 to Z=118, listed in Table~\ref{Tab2}, are chosen as the database for parameter fitting. Using the method of genetic algorithms~\cite{Car01}, a set of parameters is obtained, i.e. $a=0.5654 fm, V_0=189.53 MeV, P_{\alpha}=0.7$. And the RMS deviation is $\sqrt{\frac{\Delta}{164}}=0.350$.

\begin{figure}
\centering
\includegraphics[width=0.5\textwidth]{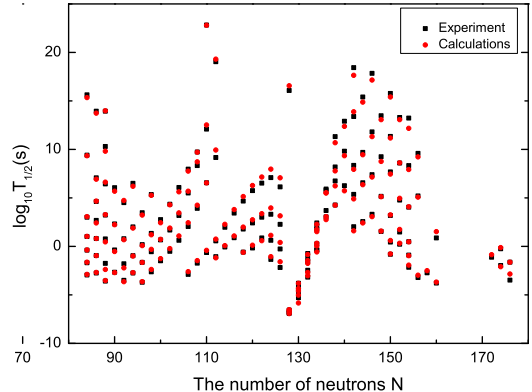}
\caption{(Color online) Logarithmic of half-lives as a function of neutron numbers of parent nuclei. The black squares and red circles represent the experimental data and calculated results, respectively.}
\label{Fig1}
\end{figure}

Systematic variations of the half-lives as a function of the neutron numbers $N$ of the parent nuclei are drawn on Fig.~\ref{Fig1}. The black squares and red circles represent experimental half-lives and calculated ones, respectively. As we can see that the theoretical results can well reproduce the experimental data of $\alpha$ decay half-lives, although the magnitude of half-lives vary in a very wide range from $10^{-7}$ s to $10^{22}$ s. This shows that our parameters are effective, and besides, the assumption that $\alpha$ particle preformation probability keeps constant is satisfactory. It noticed that $\alpha$ decay half-life $T_{1/2}$ is extremely sensitive to decay energy $Q_{\alpha}$, for example the decay energy increasing 2.6 times results in 18 order of magnitude shorter half-life for N=84 isotones. A decrease in symmetry energy may be responsible for the stability weakened~\cite{Wan14}.

\begin{figure}
\centering
\includegraphics[width=0.5\textwidth]{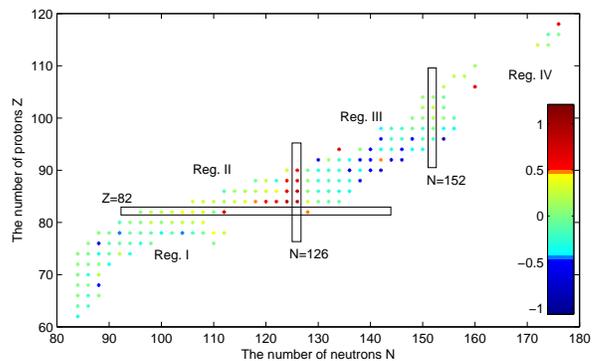}
\caption{(Color online)Logarithmic of deviation on a colour-map as a function of neutron numbers N and proton numbers Z of parent nuclei}
\label{Fig2}
\end{figure}

Furthermore it is obvious that the size of deviation between experimental half-lives and calculated results roughly increases with the increasing valence nucleon especially for nuclide with N=82-126. In order to more clearly show the results, the deviations for 164 nuclei with neutron numbers N larger than 82 are plotted as a colour-map on Fig.~\ref{Fig2}. The complete area is divided into four regions by spherical magic number Z=82, N=126 and deformed magic number N=152~\cite{Zha09}. The deviations in the area close to the magic number are greater than 1, indicating the calculated half-life is small, suggesting predicted $\alpha$ particle preformation probability in this region is too large. This rule is significant in particular for spherical magic numbers Z=82 and N=126, where clustering induced by the pairing mode is inhibited~\cite{Qi09}. In general, the predicted $\alpha$ particle preformation probabilities of nuclei in Reg. I and III are small, while the predictions in Reg. II and IV are large. In other words, the preformation probabilities in Reg. I and III are greater than those in Reg. II and IV. The reason why the preformation probabilities in Reg. II are small is that the nuclei are close to magic number Z=82 and N=126. And the reason for small preformation probabilities in Reg. IV may be the nucleus are approaching the next proton and neutron shell closure, such as the doubly magic spherical nuclei at (Z=114, N=184), (Z=120, N=172), or at (Z=126, N=184) depending on different parameters~\cite{Rut97}, and the shell effect appears again. In conclusion, the preformation probabilities vary with the distance from the magic number. 

\subsection{Hindrance factor}

Based on above methods, the calculated results are greater than the experimental data for some nuclei especially around the magic nuclei, and smaller for other nuclei. To describe the deviations in size the hindrance factor is introduced, which varies due to the nuclear shell structure.

\begin{table}
\caption{The fitness of magic number for superheavy nuclei to the $\alpha$ particle preformation probabilities}
\begin{tabular}{ccccccc}
\hline\label{Tab1}
Reg. & number & $Z_1$ & $Z_2$ & $N_1$ & $N_2$ & $\overline{\sigma^2}$ \\
\hline
I    &   61   &  50   & 82    & 82    & 126   & 0.0358\\
II   &   26   &  82   & 126   & 82    & 126   & 0.0159\\
II   &   26   &  82   & 120   & 82    & 126   & 0.0159\\
II   &   26   &  82   & 114   & 82    & 126   & 0.0160\\
III  &   59   &  82   & 126   & 126   & 152   & 0.0495\\
III  &   59   &  82   & 120   & 126   & 152   & 0.0496\\
III  &   59   &  82   & 114   & 126   & 152   & 0.0498\\
IV   &   13   &  82   & 126   & 152   & 184   & 0.0131\\
IV   &   13   &  82   & 120   & 152   & 172   & 0.0129\\
IV   &   13   &  82   & 114   & 152   & 184   & 0.0129\\
\hline \hline
\end{tabular}
\end{table}

We fit to extracted hindrance factors using the analytic expression of eq.(9), which has taken into account the nuclear shell effect and proton-neutron interaction. Z=50, 82, N=82, 126, 184 are well known magic number for neutron and proton. However, we do not know exactly the magic number for superheavy nuclei, and the predicted proton and neutron magic numbers for superheavy nuclei depend on the models and force parameters. According to the investigation within various parametrizations of relativistic and nonrelativistic nuclear mean-field models and prediction in ref.\cite{Rut97}, we fit the preformation probabilitite to the extracted ones with different protons magic number and the results are listed in table~\ref{Tab1}. The fitness is defined as $\overline{\sigma^2}=\frac{\sum_{i=1}^{n}(y_{fit}-y_{data})^2}{n}$. The fitness of (Z=114, N=184) is the worst in Reg. II and III, indicating that Z=114 is unlikely to be the next protons magic number in this region of mass, which is consistent with the result in ref.\cite{Don11}.

\begin{table}[!htb]
\caption{The parameters of hindrance factor for even-even nuclei from four different regions. Reg. I is $50 < Z \leqslant 82$ and $82 < N \leqslant 126$, Reg. II is $82 < Z \leqslant 126$ and $82 < N \leqslant 126$, Reg. III is $82\leqslant Z \leqslant 126$ and $126 < N \leqslant 152$, Reg. IV is $82 < Z \leqslant 126$ and $152 < N \leqslant 184$.}
\begin{tabular}{cccccc}
\hline\label{Tab2}
Reg. & $a$ & $b$ & $c$ & $d$ & $e$ \\
\hline
I   & 1.7828  & -0.0017 & -0.0015 & -0.0087 & 0.0011 \\
II  & 9.9252  & -0.0054 & -0.0029 & -0.0417 & 0.0033  \\
III &15.803   &  0.0012 & -0.0004 & -0.0744 & 0.0052  \\
IV & -19.5004 &   0.0042&  -0.0010&   0.0686& -0.0019 \\
\hline \hline
\end{tabular}
\end{table}

\begin{figure}[!htb]
\centering
\includegraphics[width=0.5\textwidth]{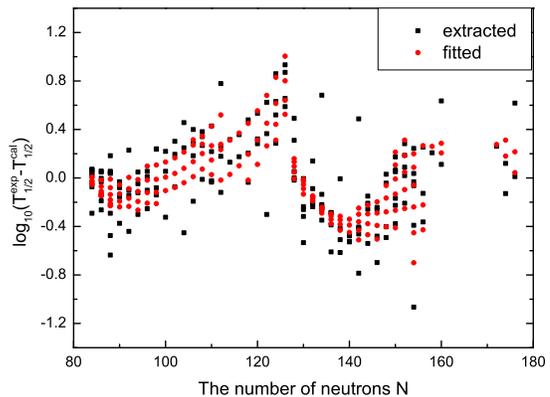}
\caption{(Color online) Logarithmic of hindrance factor as a function of neutron numbers of parent nuclei. The black squares and red circles represent the extracted data and fitted results, respectively.}
\label{Fig3}
\end{figure}

In this work, Z=126 is our choice and the obtained parameters are listed in table~\ref{Tab2}. When the hindrance factor increases, and instead the preformation probability decreases. The extracted hindrance factors and fitted ones as a function of neutron numbers $N$ are drawn on Fig.~\ref{Fig3}. As we can see the trend for fitted hindrance factors is similar to the extracted ones. Based on the improvement of hindrance factor, our results can reproduce the experimental half-lives within a factor of 2 for most nuclei, and the RMS deviation drop to $\sqrt{\frac{\Delta}{164}}=0.205$.

As can be seen from table~\ref{Tab2}, the absolute value of parameters $b$ and $c$ in Reg. II are both maximum, showing the preformation probabilities of nuclei in Reg. II are strong parabolic due to these nuclei sandwiched between magic numbers Z=82 and N=126. And it is interested that parameter $d$ is negatively correlated with parameter $a$ and $e$. When the term of nuclei mass $A$ makes greater contributions to the hindrance factors in different regions the contributions from constant term and integrated valence neutron-proton term are smaller, which shows that these three terms are related. A physical explanation of this relationship closely associated with the nucleons clustering in heavy nuclei is worth exploring in the future.

\clearpage
\begingroup
\renewcommand*{\arraystretch}{1.3}
\begin{longtable*}{ccccccc}
\caption{Comparisons among experimental $\alpha$ decay half-lives $T^{exp}_{1/2}$ and calculated results within our work $T^{cal}_{1/2}$, generalized liquid drop model $T^{GLDM}_{1/2}$~\cite{Bao14}, density-dependent cluster model $T^{DDCM}_{1/2}$~\cite{Xu08,Xu05}  and $T^{sym}_{1/2}$ taking into account isospin-dependent nuclear potential in eq.(11) of even-even nuclei with proton number Z=62-118.}\label{Tab3} \\
\hline Nuclei & N & $T^{exp}_{1/2}(s)$ & $T^{cal}_{1/2}(s)$ & $T^{GLDM}_{1/2}(s)$ & $T^{DDCM}_{1/2}(s)$ & $T^{sym}_{1/2}(s)$ \\ \hline
\endfirsthead
\multicolumn{7}{c}%
{{\tablename\ \thetable{} -- continued from previous page}} \\
\hline Nuclei & N & $T^{exp}_{1/2}(s)$ & $T^{cal}_{1/2}(s)$ & $T^{GLDM}_{1/2}(s)$ & $T^{DDCM}_{1/2}(s)$ & $T^{sym}_{1/2}(s)$ \\ \hline
\endhead
\hline \multicolumn{7}{r}{{Continued on next page}} \\
\endfoot
\hline \hline
\endlastfoot
$^{146}$Sm&84 &$2.2\times10^{15}$&$ 4.3\times10^{15}$&$ 13 \times10^{15}$&$2.3\times10^{15}$&$4.0\times10^{15}$\\
$^{148}$Gd&84 &$2.2\times10^{9 }$&$2.4 \times10^{ 9}$&$ 6.4\times10^{ 9}$&$1.7\times10^{9}$ &$2.5\times10^{9 }$\\
$^{150}$Gd&86 &$5.6\times10^{13}$&$ 6.4\times10^{13}$&$ 22 \times10^{13}$&$6.0\times10^{13}$&$6.7\times10^{13}$\\
$^{150}$Dy&84 &$1.2\times10^{3 }$&$0.91\times10^{ 3}$&$ 2.4\times10^{ 3}$&$0.83\times10^{3}$&$1.0\times10^{3 }$\\
$^{152}$Dy&86 &$8.6\times10^{6 }$&$8.6 \times10^{ 6}$&$ 25 \times10^{6 }$&$8.5\times10^{6}$ &$9.2\times10^{6 }$\\
$^{154}$Dy&88 &$9.5\times10^{13}$&$ 5.1\times10^{13}$&$ 20 \times10^{13}$&$5.1\times10^{13}$&$5.4\times10^{13}$\\
$^{152}$Er&84 &$1.1\times10^{1 }$&$0.9 \times10^{ 1}$&$ 2.1\times10^{ 1}$&$0.86\times10^{1}$&$1.1\times10^{1 }$\\
$^{154}$Er&86 &$4.8\times10^{4 }$&$3.1 \times10^{ 4}$&$ 7.4\times10^{ 4}$&$3.1\times10^{4}$ &$2.8\times10^{4 }$\\
$^{156}$Er&88 &$6.7\times10^{9 }$&$12  \times10^{9 }$&$39  \times10^{9 }$&$-$               &$15 \times10^{9 }$\\
$^{154}$Yb&84 &$4.4\times10^{-1}$&$ 3.6\times10^{-1}$&$ 6.9\times10^{-1}$&$3.4\times10^{-1}$&$3.6\times10^{-1}$\\
$^{156}$Yb&86 &$2.6\times10^{2 }$&$3.5 \times10^{ 2}$&$ 8.7\times10^{2 }$&$4.0\times10^{2}$ &$3.6\times10^{2 }$\\
$^{158}$Yb&88 &$4.3\times10^{6 }$&$1.7 \times10^{ 6}$&$ 4.5\times10^{6 }$&$2.0\times10^{6}$ &$1.7\times10^{6 }$\\
$^{156}$Hf&84 &$2.4\times10^{-2}$&$ 1.9\times10^{-2}$&$ 3.5\times10^{-2}$&$1.5\times10^{-2}$&$2.2\times10^{-2}$\\
$^{158}$Hf&86 &$6.4\times10^{0 }$&$5   \times10^{0 }$&$11  \times10^{0 }$&$5.6\times10^{0}$ &$5.4\times10^{0 }$\\
$^{160}$Hf&88 &$1.9\times10^{3 }$&$1.1 \times10^{ 3}$&$ 2.7\times10^{3 }$&$1.6\times10^{3}$ &$1.3\times10^{3 }$\\
$^{162}$Hf&90 &$4.9\times10^{5 }$&$6.7 \times10^{ 5}$&$ 15 \times10^{5 }$&$8.2\times10^{5}$ &$6.4\times10^{5 }$\\
$^{158}$W&84  &$1.3\times10^{-3}$&$ 1.1\times10^{-3}$&$ 2  \times10^{-3}$&$0.9\times10^{-3}$&$1.3\times10^{-3}$\\
$^{160}$W&86  &$1.1\times10^{-1}$&$ 1  \times10^{-1}$&$1.6\times10^{-1}$&$0.82\times10^{-1}$&$0.9\times10^{-1}$\\
$^{162}$W&88  &$3  \times10^{0 }$&$4.3 \times10^{ 0}$&$ 5.9\times10^{0 }$&$3.3\times10^{0}$ &$2.6\times10^{0 }$\\
$^{164}$W&90  &$1.7\times10^{2 }$&$1.3 \times10^{ 2}$&$ 3.1\times10^{2 }$&$2.0\times10^{2}$ &$1.9\times10^{2 }$\\
$^{166}$W&92  &$5.5\times10^{4 }$&$1.9 \times10^{ 4}$&$ 4.1\times10^{4 }$&$2.8\times10^{4}$ &$2.1\times10^{4 }$\\
$^{168}$W&94  &$1.6\times10^{6 }$&$1.7 \times10^{ 6}$&$ 4.6\times10^{6 }$&$-$               &$2.2\times10^{6 }$\\
$^{162}$Os&86 &$2.1\times10^{-3}$&$ 1.8\times10^{-3}$&$ 3.6\times10^{-3}$&$1.8\times10^{-3}$&$2.0\times10^{-3}$\\
$^{164}$Os&88 &$4.1\times10^{-3}$&$ 15 \times10^{-3}$&$ 31 \times10^{-3}$&$1.8\times10^{-3}$&$18 \times10^{-3}$\\
$^{166}$Os&90 &$3  \times10^{-1}$&$ 3.2\times10^{-1}$&$ 4.9\times10^{-1}$&$3.7\times10^{-1}$&$2.7\times10^{-1}$\\
$^{168}$Os&92 &$4.9\times10^{0 }$&$4.2 \times10^{ 0}$&$ 9.1\times10^{0 }$&$6.5\times10^{0}$ &$4.7\times10^{0 }$\\
$^{170}$Os&94 &$7.8\times10^{1 }$&$6.6 \times10^{ 1}$&$ 14 \times10^{1 }$&$10\times10^{1}$  &$7.1\times10^{1 }$\\
$^{172}$Os&96 &$1.7\times10^{3 }$&$1.9 \times10^{ 3}$&$ 3.9\times10^{ 3}$&$3.3\times10^{3}$ &$2.2\times10^{3 }$\\
$^{174}$Os&98 &$1.8\times10^{5 }$&$1.3 \times10^{ 5}$&$ 2.8\times10^{ 5}$&$2.4\times10^{5}$ &$1.7\times10^{5 }$\\
$^{186}$Os&110&$6.3\times10^{22}$&$ 7.3\times10^{22}$&$ 10 \times10^{22}$&$4.2\times10^{22}$&$6.6\times10^{22}$\\
$^{166}$Pt&88 &$3.0\times10^{-4}$&$2.6 \times10^{-4}$&$4.5 \times10^{-4}$&$2.3\times10^{-4}$&$3.0\times10^{-4}$\\
$^{168}$Pt&90 &$2.0\times10^{-3}$&$1.9 \times10^{-3}$&$3.3 \times10^{-3}$&$1.8\times10^{-3}$&$2.1\times10^{-3}$\\
$^{170}$Pt&92 &$6.0\times10^{-3}$&$14  \times10^{-3}$&$26  \times10^{-3}$&$18 \times10^{-3}$&$15 \times10^{-3}$\\
$^{172}$Pt&94 &$1.0\times10^{-1}$&$0.86\times10^{-1}$&$1.6 \times10^{-1}$&$1.2\times10^{-1}$&$1.0\times10^{-1}$\\
$^{174}$Pt&96 &$1.2\times10^{0 }$&$1.1 \times10^{0 }$&$1.7 \times10^{0 }$&$1.5\times10^{0 }$&$1.1\times10^{0 }$\\
$^{176}$Pt&98 &$1.6\times10^{1 }$&$1.7 \times10^{1 }$&$2.6 \times10^{1 }$&$2.5\times10^{1 }$&$1.9\times10^{1 }$\\
$^{178}$Pt&100&$2.7\times10^{2 }$&$4.5 \times10^{2 }$&$5.8 \times10^{2 }$&$2.5\times10^{2 }$&$6.5\times10^{2 }$\\
$^{180}$Pt&102&$1.9\times10^{4 }$&$1.9 \times10^{4 }$&$2.3 \times10^{4 }$&$2.7\times10^{4 }$&$2.0\times10^{4 }$\\
$^{182}$Pt&104&$4.2\times10^{5 }$&$11  \times10^{5 }$&$7.9 \times10^{5 }$&$9.4\times10^{5 }$&$7.6\times10^{5 }$\\

$^{184}$Pt&106&$5.9\times10^{7 }$&$9.4 \times10^{7 }$&$10  \times10^{7 }$&$12 \times10^{7 }$&$10 \times10^{7 }$\\
$^{186}$Pt&108&$5.3\times10^{9 }$&$6.4 \times10^{9 }$&$7.1 \times10^{9 }$&$8.6\times10^{9 }$&$7.8\times10^{9 }$\\
$^{188}$Pt&110&$3.3\times10^{12}$&$1.7 \times10^{12}$&$1.6 \times10^{12}$&$1.7\times10^{12}$&$2.2\times10^{12}$\\
$^{190}$Pt&112&$2.0\times10^{19}$&$1.9 \times10^{19}$&$1.8 \times10^{19}$&$1.7\times10^{19}$&$2.5\times10^{19}$\\
$^{172}$Hg&92 &$2.3\times10^{-4}$&$3.2 \times10^{-4}$&$3.9 \times10^{-4}$&$2.4\times10^{-4}$&$2.8\times10^{-4}$\\
$^{174}$Hg&94 &$2.0\times10^{-3}$&$1.7 \times10^{-3}$&$2.6 \times10^{-3}$&$1.9\times10^{-3}$&$1.9\times10^{-3}$\\
$^{176}$Hg&96 &$2.3\times10^{-2}$&$2.1 \times10^{-2}$&$2.9 \times10^{-2}$&$2.4\times10^{-2}$&$2.4\times10^{-2}$\\
$^{178}$Hg&98 &$5.0\times10^{-1}$&$2.8 \times10^{-1}$&$3.5 \times10^{-1}$&$3.2\times10^{-1}$&$3.1\times10^{-1}$\\
$^{180}$Hg&100&$5.4\times10^{0 }$&$4.8 \times10^{0 }$&$5.2 \times10^{0 }$&$5.5\times10^{0 }$&$5.4\times10^{0 }$\\
$^{182}$Hg&102&$7.8\times10^{1 }$&$5.4 \times10^{1 }$&$5.6 \times10^{1 }$&$6.5\times10^{1 }$&$7.6\times10^{1 }$\\
$^{184}$Hg&104&$2.8\times10^{3 }$&$1.9 \times10^{3 }$&$1.6 \times10^{3 }$&$2.1\times10^{3 }$&$2.2\times10^{3 }$\\
$^{186}$Hg&106&$5.0\times10^{5 }$&$4.5 \times10^{5 }$&$3.0 \times10^{5 }$&$4.2\times10^{5 }$&$4.6\times10^{5 }$\\
$^{188}$Hg&108&$5.2\times10^{8 }$&$3.4 \times10^{8 }$&$2.4 \times10^{8 }$&$3.4\times10^{8 }$&$7.8\times10^{8 }$\\
$^{178}$Pb&96 &$2.3\times10^{-4}$&$2.5 \times10^{-4}$&$3.0 \times10^{-4}$&$-$               &$2.8\times10^{-4}$\\
$^{180}$Pb&98 &$4.2\times10^{-3}$&$3.1 \times10^{-3}$&$3.2 \times10^{-3}$&$-$               &$4.2\times10^{-3}$\\
$^{182}$Pb&100&$5.5\times10^{-2}$&$4.5 \times10^{-2}$&$3.9 \times10^{-2}$&$5.7\times10^{-2}$&$4.8\times10^{-2}$\\
$^{184}$Pb&102&$6.1\times10^{-1}$&$4.5 \times10^{-1}$&$3.6 \times10^{-1}$&$4.0\times10^{-1}$&$5.3\times10^{-1}$\\
$^{186}$Pb&104&$1.2\times10^{1 }$&$0.69\times10^{1 }$&$4.4 \times10^{1 }$&$0.52\times10^{1}$&$0.8\times10^{1 }$\\
$^{188}$Pb&106&$2.8\times10^{2 }$&$2.1 \times10^{2 }$&$1.1 \times10^{2 }$&$1.5\times10^{2}$ &$2.9\times10^{2 }$\\
$^{190}$Pb&108&$1.8\times10^{4 }$&$1.8 \times10^{4 }$&$0.76\times10^{4 }$&$1.0\times10^{4}$ &$2.0\times10^{4 }$\\
$^{192}$Pb&110&$3.5\times10^{6 }$&$9.8 \times10^{6 }$&$2.0 \times10^{6 }$&$2.8\times10^{6}$ &$7.0\times10^{6 }$\\
$^{194}$Pb&112&$8.8\times10^{9 }$&$4.9 \times10^{9 }$&$1.4 \times10^{9 }$&$1.9\times10^{9}$ &$6.2\times10^{9 }$\\
$^{210}$Pb&128&$3.7\times10^{16}$&$1.7 \times10^{16}$&$2.0 \times10^{16}$&$1.1\times10^{16}$&$2.2\times10^{16}$\\
$^{190}$Po&106&$2.5\times10^{-3}$&$2.7\times10^{-3}$&$1.7\times10^{-3}$&$1.9\times10^{-3} $ &$2.7\times10^{-3}$\\
$^{192}$Po&108&$3.4\times10^{-2}$&$3.5\times10^{-2}$&$2.2\times10^{-2}$&$2.8 \times10^{-2}$ &$3.1\times10^{-2}$\\
$^{194}$Po&110&$3.9\times10^{-1}$&$4.4\times10^{-1}$&$2.5\times10^{-1}$&$3.9 \times10^{-1}$ &$4.2\times10^{-1}$\\
$^{196}$Po&112&$5.7\times10^{0} $&$7.2\times10^{0}$&$3.6\times10^{0}$ &$6.3 \times10^{0} $  &$6.5\times10^{0} $\\
$^{198}$Po&114&$1.9\times10^{2} $&$1.9\times10^{2}$&$0.81\times10^{2}$&$1.6 \times10^{2} $  &$1.8\times10^{2} $\\
$^{200}$Po&116&$6.2\times10^{3} $&$6.4\times10^{3}$&$2.0\times10^{3}$ &$4.2 \times10^{3} $  &$6.9\times10^{3} $\\
$^{202}$Po&118&$1.4\times10^{5} $&$1.3\times10^{5}$&$0.38\times10^{5}$&$0.87\times10^{5} $  &$1.2\times10^{5} $\\
$^{204}$Po&120&$1.9\times10^{6} $&$2.0\times10^{6}$&$0.45\times10^{6}$&$1.1 \times10^{6} $  &$2.1\times10^{6} $\\
$^{206}$Po&122&$1.4\times10^{7} $&$1.6\times10^{7}$&$0.30\times10^{7}$&$0.70\times10^{7} $  &$1.8\times10^{7} $\\
$^{208}$Po&124&$9.1\times10^{7} $&$8.5\times10^{7}$&$0.12\times10^{7}$&$2.8 \times10^{7} $  &$9.2\times10^{7} $\\
$^{210}$Po&126&$1.2\times10^{7} $&$1.4\times10^{7}$&$0.092\times10^{7}$&$0.23\times10^{7}$  &$1.2\times10^{7} $\\
$^{212}$Po&128&$3.0\times10^{-7}$&$2.0\times10^{-7}$&$2.5\times10^{-7}$&$2.3\times10^{-7}$  &$2.6\times10^{-7}$\\
$^{214}$Po&130&$1.6\times10^{-4}$&$1.6\times10^{-4}$&$1.6\times10^{-4}$&$2.3\times10^{-4}$  &$1.9\times10^{-4}$\\
$^{216}$Po&132&$1.5\times10^{-1}$&$1.3\times10^{-1}$&$1.5\times10^{-1}$&$2.4\times10^{-1}$  &$1.3\times10^{-1}$\\
$^{218}$Po&134&$1.9\times10^{2} $&$1.3\times10^{2}$ &$2.1\times10^{2}$&$3.5\times10^{2}$    &$1.4\times10^{2} $\\
$^{198}$Rn&112&$6.6\times10^{-2}$&$8.3\times10^{-2}$&$8.1\times10^{-2}$&$-$                 &$9.9\times10^{-2}$\\
$^{200}$Rn&114&$1.2\times10^{0} $&$0.95\times10^{0}$&$0.83\times10^{0} $&$1.7\times10^{0}$  &$0.9\times10^{0} $\\
$^{202}$Rn&116&$1.2\times10^{1} $&$1.0\times10^{1} $&$0.71\times10^{1} $&$1.8\times10^{1}$  &$1.0\times10^{1} $\\
$^{204}$Rn&118&$1.0\times10^{2} $&$0.97\times10^{2}$&$0.49\times10^{2} $&$1.3\times10^{2} $ &$1.0\times10^{2} $\\
$^{206}$Rn&120&$5.5\times10^{2} $&$5.4\times10^{2} $&$2.1 \times10^{2} $&$6.0\times10^{2} $ &$6.1\times10^{2} $\\
$^{208}$Rn&122&$2.4\times10^{3} $&$2.9\times10^{3} $&$0.65\times10^{3} $&$1.9\times10^{3} $ &$2.3\times10^{3} $\\

$^{210}$Rn&124&$9.0\times10^{3} $&$8.7\times10^{3} $&$1.6 \times10^{3} $&$5.0\times10^{3} $  &$9.2\times10^{3} $\\
$^{212}$Rn&126&$1.4\times10^{3} $&$1.2\times10^{3} $&$0.15\times10^{3} $&$0.50\times10^{3}$  &$1.2\times10^{3} $\\
$^{214}$Rn&128&$2.7\times10^{-7}$&$3.3\times10^{-7}$&$2.9 \times10^{-7}$&$2.8\times10^{-7}$  &$2.3\times10^{-7}$\\
$^{216}$Rn&130&$4.5\times10^{-5}$&$7.3\times10^{-5}$&$8.2 \times10^{-5}$&$11\times10^{-5} $  &$7.5\times10^{-5}$\\
$^{218}$Rn&132&$3.5\times10^{-2}$&$4.3\times10^{-2}$&$5.3 \times10^{-2}$&$9.2\times10^{-2}$  &$4.8\times10^{-2}$\\
$^{220}$Rn&134&$5.6\times10^{1} $&$6.8\times10^{1} $&$8.8 \times10^{1} $&$17\times10^{1}  $  &$6.5\times10^{1} $\\
$^{222}$Rn&136&$3.3\times10^{5} $&$3.4\times10^{5} $&$6.3 \times10^{5} $&$10\times10^{5}  $  &$3.0\times10^{5} $\\
$^{206}$Ra&118&$2.4\times10^{-1}$&$2.5 \times10^{-1}$&$2.1 \times10^{-1}$&$4.6\times10^{-1}$ &$3.5\times10^{-1}$\\
$^{208}$Ra&120&$1.3\times10^{0 }$&$0.88\times10^{0 }$&$0.57\times10^{0 }$&$1.4\times10^{0 }$ &$0.9\times10^{0 }$\\
$^{210}$Ra&122&$3.8\times10^{0 }$&$12  \times10^{0 }$&$1.3 \times10^{0 }$&$3.3\times10^{0 }$ &$4.0\times10^{0 }$\\
$^{212}$Ra&124&$1.4\times10^{1 }$&$1.2 \times10^{1 }$&$0.33\times10^{1 }$&$9.2\times10^{1 }$ &$1.4\times10^{1 }$\\
$^{214}$Ra&126&$2.5\times10^{0 }$&$2.4 \times10^{0 }$&$0.41\times10^{0 }$&$1.3\times10^{0 }$ &$2.9\times10^{0 }$\\
$^{216}$Ra&128&$1.8\times10^{-7}$&$1.8 \times10^{-7}$&$2.5 \times10^{-7}$&$2.5\times10^{-7}$ &$1.6\times10^{-7}$\\
$^{218}$Ra&130&$2.5\times10^{-5}$&$4.0 \times10^{-5}$&$4.7 \times10^{-5}$&$6.7\times10^{-5}$ &$3.6\times10^{-5}$\\
$^{220}$Ra&132&$1.8\times10^{-2}$&$2.0 \times10^{-2}$&$2.5 \times10^{-2}$&$4.8\times10^{-2}$ &$2.2\times10^{-2}$\\
$^{222}$Ra&134&$3.4\times10^{1 }$&$3.5 \times10^{1 }$&$4.5 \times10^{1 }$&$10 \times10^{1 }$ &$3.5\times10^{1 }$\\
$^{224}$Ra&136&$3.2\times10^{5 }$&$2.8 \times10^{5 }$&$5.1 \times10^{5 }$&$7.7\times10^{5 }$ &$2.9\times10^{5 }$\\
$^{226}$Ra&138&$5.0\times10^{10}$&$7.6 \times10^{10}$&$15  \times10^{10}$&$12.4\times10^{10}$&$4.5\times10^{10}$\\
$^{214}$Th&124&$8.7\times10^{-2}$&$9.2 \times10^{-2}$&$0.04\times10^{-2}$&$10\times10^{-2}$  &$9.9\times10^{-2}$\\
$^{216}$Th&126&$2.6\times10^{-2}$&$2.2 \times10^{-2}$&$0.62\times10^{-2}$&$-$  &$2.4\times10^{-2}$\\
$^{218}$Th&128&$1.2\times10^{-7}$&$1.2 \times10^{-7}$&$2.2 \times10^{-7}$&$1.3\times10^{-7}$ &$1.2\times10^{-7}$\\
$^{220}$Th&130&$9.7\times10^{-6}$&$16  \times10^{-6}$&$20  \times10^{-6}$&$18\times10^{-6}$  &$13 \times10^{-6}$\\
$^{222}$Th&132&$2.1\times10^{-3}$&$2.4 \times10^{-3}$&$3.0 \times10^{-3}$&$3.5\times10^{-3}$ &$2.2\times10^{-3}$\\
$^{224}$Th&134&$1.1\times10^{0 }$&$1.0 \times10^{0 }$&$1.3 \times10^{0 }$&$1.4\times10^{0}$  &$1.0\times10^{0 }$\\
$^{226}$Th&136&$1.2\times10^{3 }$&$2.4 \times10^{3 }$&$2.8 \times10^{3 }$&$2.1\times10^{3}$  &$2.3\times10^{3 }$\\
$^{228}$Th&138&$6.0\times10^{7 }$&$6.5 \times10^{7 }$&$13  \times10^{7 }$&$7.6\times10^{7}$  &$6.7\times10^{7 }$\\
$^{230}$Th&140&$2.4\times10^{12}$&$2.9 \times10^{12}$&$8.7 \times10^{12}$&$3.0\times10^{12}$ &$2.8\times10^{12}$\\
$^{232}$Th&142&$4.4\times10^{17}$&$8.3 \times10^{17}$&$33  \times10^{17}$&$6.5\times10^{17}$ &$6.3\times10^{17}$\\
$^{222}$U&130 &$1.5\times10^{-6}$&$ 3.8\times10^{-6}$&$ 6.3\times10^{-6}$&$3.7\times10^{-6}$ &$3.1\times10^{-6}$\\
$^{224}$U&132 &$9.4\times10^{-4}$&$ 4.4\times10^{-4}$&$ 6.2\times10^{-4}$&$8.0\times10^{-4}$ &$3.7\times10^{-4}$\\
$^{226}$U&134 &$2.7\times10^{-1}$&$ 2.6\times10^{-1}$&$ 3.2\times10^{-1}$&$3.6\times10^{-1}$ &$2.6\times10^{-1}$\\
$^{228}$U&136 &$5.7\times10^{2 }$&$5.6 \times10^{2 }$&$6.1 \times10^{2 }$&$5.2\times10^{2}$  &$4.6\times10^{2 }$\\
$^{230}$U&138 &$1.7\times10^{6 }$&$2.5 \times10^{6 }$&$3.1 \times10^{6 }$&$4.3\times10^{6}$  &$2.0\times10^{6 }$\\
$^{232}$U&140 &$2.2\times10^{9 }$&$2.7 \times10^{9 }$&$5.1 \times10^{9 }$&$5.1\times10^{9}$  &$3.6\times10^{9 }$\\
$^{234}$U&142 &$7.7\times10^{12}$&$0.93\times10^{12}$&$ 2.4\times10^{12}$&$14\times10^{12}$  &$-$               \\
$^{236}$U&144 &$7.4\times10^{14}$&$8.7 \times10^{14}$&$ 32 \times10^{14}$&$14\times10^{14}$  &$9.8\times10^{14}$\\
$^{238}$U&146 &$1.4\times10^{17}$&$2.2 \times10^{17}$&$ 11 \times10^{17}$&$3.3\times10^{17}$ &$2.1\times10^{17}$\\

$^{228}$Pu&134&$2.1\times10^{ 0}$&$0.24\times10^{ 0}$&$0.29\times10^{ 0}$&$-$                &$-$               \\
$^{232}$Pu&138&$1.8\times10^{ 4}$&$0.88\times10^{ 4}$&$0.91\times10^{ 4}$&$1.2\times10^{4}$  &$0.7\times10^{ 4}$\\
$^{234}$Pu&140&$5.3\times10^{ 5}$&$8.2 \times10^{5 }$&$6.1 \times10^{5 }$&$6.2\times10^{5}$  &$4.4\times10^{ 5}$\\
$^{236}$Pu&142&$9  \times10^{ 7}$&$10  \times10^{7 }$&$11  \times10^{7 }$&$12\times10^{7}$   &$7.0\times10^{ 7}$\\
$^{238}$Pu&144&$2.8\times10^{ 9}$&$2.1 \times10^{9 }$&$3.7 \times10^{9 }$&$3.4\times10^{9}$  &$2.1\times10^{ 9}$\\
$^{240}$Pu&146&$2.1\times10^{11}$&$ 2.6\times10^{11}$&$ 4.7\times10^{11}$&$3.2\times10^{11}$ &$2.1\times10^{11}$\\
$^{242}$Pu&148&$1.2\times10^{13}$&$ 1.2\times10^{13}$&$ 3.2\times10^{13}$&$1.8\times10^{13}$ &$1.3\times10^{13}$\\
$^{244}$Pu&150&$2.5\times10^{15}$&$ 2.3\times10^{15}$&$ 7.5\times10^{15}$&$2.9\times10^{15}$ &$2.5\times10^{15}$\\
$^{238}$Cm&142&$7.9\times10^{4 }$&$11  \times10^{4 }$&$8.5 \times10^{4 }$&$-$                &$8.1\times10^{4 }$\\
$^{240}$Cm&144&$2.3\times10^{6 }$&$1.9 \times10^{6 }$&$1.5 \times10^{6 }$&$2.1\times10^{6}$  &$1.5\times10^{6 }$\\
$^{242}$Cm&146&$1.4\times10^{7 }$&$1.3 \times10^{7 }$&$1.1 \times10^{7 }$&$1.5\times10^{7}$  &$1.4\times10^{7 }$\\
$^{244}$Cm&148&$5.7\times10^{8 }$&$9.3 \times10^{8 }$&$5.2 \times10^{8 }$&$5.4\times10^{8}$  &$4.6\times10^{8 }$\\
$^{246}$Cm&150&$1.5\times10^{11}$&$ 1.2\times10^{11}$&$ 1.8\times10^{11}$&$1.3\times10^{11}$ &$1.2\times10^{11}$\\
$^{248}$Cm&152&$1.2\times10^{13}$&$ 1.1\times10^{13}$&$ 2  \times10^{13}$&$1.1\times10^{13}$ &$1.2\times10^{13}$\\
$^{250}$Cm&154&$1.5\times10^{12}$&$ 3.5\times10^{12}$&$ 18 \times10^{12}$&$8.6\times10^{12}$ &$3.5\times10^{12}$\\
$^{240}$Cf&142&$4.1\times10^{1 }$&$6   \times10^{1 }$&$3.5 \times10^{1 }$&$-$                &$4.3\times10^{1 }$\\
$^{242}$Cf&144&$2.6\times10^{2 }$&$2.2 \times10^{2 }$&$1.7 \times10^{2 }$&$3.0\times10^{2}$  &$2.9\times10^{2 }$\\
$^{244}$Cf&146&$1.2\times10^{3 }$&$1.3 \times10^{3 }$&$0.82\times10^{ 3}$&$1.4\times10^{3}$  &$1.3\times10^{3 }$\\
$^{246}$Cf&148&$1.3\times10^{5 }$&$0.98\times10^{ 5}$&$0.64\times10^{ 5}$&$1.1\times10^{5}$  &$1.3\times10^{5 }$\\
$^{248}$Cf&150&$2.9\times10^{7 }$&$3.6 \times10^{7 }$&$1.4 \times10^{7 }$&$1.9\times10^{7}$  &$2.4\times10^{7 }$\\
$^{250}$Cf&152&$4.1\times10^{8 }$&$3.1 \times10^{8 }$&$2.3 \times10^{8 }$&$2.5\times10^{8}$  &$4.4\times10^{8 }$\\
$^{252}$Cf&154&$8.6\times10^{7 }$&$7.5 \times10^{7 }$&$7.3 \times10^{7 }$&$8.3\times10^{7}$  &$8.1\times10^{7 }$\\
$^{254}$Cf&156&$1.7\times10^{9 }$&$1.5 \times10^{9 }$&$2.9 \times10^{9 }$&$2.8\times10^{9}$  &$2.0\times10^{9 }$\\
$^{248}$Fm&148&$3.8\times10^{1 }$&$3.1 \times10^{1 }$&$1.5 \times10^{1 }$&$2.5\times10^{1}$  &$3.4\times10^{1 }$\\
$^{250}$Fm&150&$1.8\times10^{3 }$&$1.5 \times10^{3 }$&$0.54\times10^{ 3}$&$1.0\times10^{3}$  &$1.9\times10^{3 }$\\
$^{252}$Fm&152&$9.1\times10^{4 }$&$6.6 \times10^{4 }$&$2.1 \times10^{4 }$&$3.7\times10^{4}$  &$8.3\times10^{4 }$\\
$^{254}$Fm&154&$1.2\times10^{4 }$&$0.63\times10^{ 4}$&$0.42\times10^{ 4}$&$0.74\times10^{4}$ &$0.6\times10^{4 }$\\
$^{256}$Fm&156&$1.2\times10^{5 }$&$0.96\times10^{ 5}$&$ 0.6\times10^{ 5}$&$1.1\times10^{5}$  &$1.2\times10^{5 }$\\
$^{252}$No&150&$4.1\times10^{0 }$&$3.5 \times10^{0 }$&$1.1 \times10^{0 }$&$2.1\times10^{0}$  &$4.5\times10^{0 }$\\
$^{254}$No&152&$5.7\times10^{1 }$&$4.6 \times10^{1 }$&$1.1 \times10^{1 }$&$2.3\times10^{1}$  &$5.7\times10^{1 }$\\
$^{256}$No&154&$2.9\times10^{0 }$&$2.7 \times10^{0 }$&$0.7 \times10^{0 }$&$1.3\times10^{0}$  &$1.6\times10^{0 }$\\
$^{254}$Rf&150&$1.7\times10^{-1}$&$ 2.5\times10^{-1}$&$ 0.6\times10^{-1}$&$0.35\times10^{-1}$&$2.9\times10^{-1}$\\
$^{256}$Rf&152&$2.1\times10^{0 }$&$3.5 \times10^{0 }$&$0.34\times10^{0 }$&$0.68\times10^{0}$ &$2.6\times10^{0 }$\\
$^{258}$Rf&154&$1.1\times10^{-1}$&$ 1.6\times10^{-1}$&$0.55\times10^{-1}$&$0.65\times10^{-1}$&$2.2\times10^{-1}$\\
$^{260}$Sg&154&$1.2\times10^{-2}$&$ 1.1\times10^{-2}$&$0.29\times10^{-2}$&$0.45\times10^{-2}$&$1.4\times10^{-2}$\\
$^{266}$Sg&160&$3.3\times10^{1 }$&$1.2 \times10^{1 }$&$0.15\times10^{1 }$&$0.82\times10^{1}$ &$1.3\times10^{1 }$\\
$^{264}$Hs&156&$1.1\times10^{-3}$&$ 1.1\times10^{-3}$&$0.25\times10^{-3}$&$0.54\times10^{-3}$&$1.3\times10^{-3}$\\
$^{266}$Hs&158&$3.1\times10^{-3}$&$ 3.5\times10^{-3}$&$0.86\times10^{-3}$&$1.8\times10^{-3}$ &$4.1\times10^{-3}$\\
$^{270}$Ds&160&$2.1\times10^{-4}$&$ 3.1\times10^{-4}$&$0.57\times10^{-4}$&$0.64\times10^{-4}$&$2.8\times10^{-4}$\\
$^{286}$Fl&172&$1.4\times10^{-1}$&$ 1.4\times10^{-1}$&$0.24\times10^{-1}$&$3.3\times10^{-1}$ &$1.6\times10^{-1}$\\
$^{288}$Fl&174&$7.5\times10^{-1}$&$ 12 \times10^{-1}$&$1.2 \times10^{-1}$&$16\times10^{-1}$  &$13 \times10^{-1}$\\
$^{290}$Lv&174&$8  \times10^{-3}$&$16  \times10^{-3}$&$2.7 \times10^{-3}$&$26\times10^{-3}$  &$12 \times10^{-3}$\\
$^{292}$Lv&176&$2.4\times10^{-2}$&$ 3.8\times10^{-2}$&$0.79\times10^{-2}$&$7.6\times10^{-2}$ &$4.0\times10^{-2}$\\
$^{294}$118&176&$1.4\times10^{-3}$&$0.37\times10^{-3}$&$0.14\times10^{-3}$&$1.2\times10^{-3}$&$0.4\times10^{-3}$\\

\end{longtable*}
\endgroup

In Table~\ref{Tab3}, we list the experimental half-lives and calculated results of even-even nuclei with proton number Z=62-118. The first and second columns denote the parent nucleus and their neutron numbers $N$, respectively. The third column is the experimental half-lives of $\alpha$ decay in unit of second. The next three columns are calculated half-lives within our work, the generalized liquid droplet model and the density-dependent cluster model, respectively. The lastest column is the calculated results of our work taking into account isospin-dependent nuclear potential of eq.(11). In general, our results are better than others, especially for the nuclei around the shell closure, with the help of analytic expression for hindrance factors. For example, the Radon isotopes stride across the neutron shell in N=126, and the preformation probability will abruptly decreases. It is shown that our results overcome the shortcoming of the shorter predicted half-life of $^{212}$Rn and $^{210}$Rn.

\begin{figure}
\centering
\includegraphics[width=.5\textwidth]{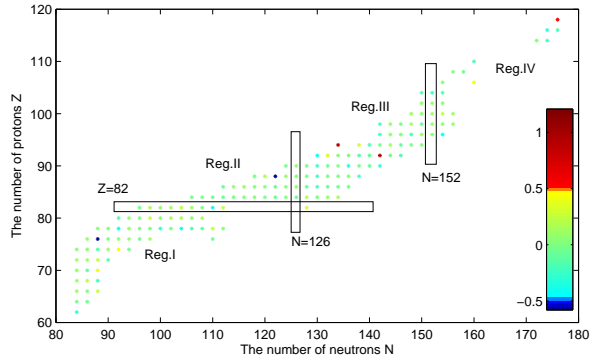}
\caption{(Color online)Logarithmic of deviation on a colour-map as a function of neutron numbers N and proton number Z of parent nuclei considering hindrance factor}
\label{Fig4}
\end{figure}

To evaluate the role of hindrance factors, we plot the deviations between the calculated and experimental half-lives again in Fig.~\ref{Fig4}. Compared with Fig.~\ref{Fig2}, the abrupt increasing of the deviations close to the magic number has been got over. 

\subsection{Isospin effect of phenomenological nuclear potential}

From the term of symmetry energy in Bethe-Weizs\"{a}cker mass formula~\cite{Wan10} to asymmetry-dependent components in nucleon-nucleon optical model physics~\cite{Kon03}, they both show that isospin effect play a role in nuclear potential. The excited analogue state in nuclei and exotic phenomenon of neutron-proton pairing could be also isospin related~\cite{War06}. Neutron and proton been treated as being different charge states of the same particles, but the fact that the strong interaction is independent on charge, lead to confusedness of origin and uncertainty of the isospin effect~\cite{Li08,Ste05}.

The accurate calculations of $\alpha$ decay half-lives has been given within two potential approach and correction of hindrance factor. Now we introduce two extra parameters to indicate the isospin effect of nuclear potential. The new depth $V_0$ and diffuseness $a$ of nuclear potential could be given as
\begin{eqnarray}\label{12}
V_0&=193.57-75.61\frac{N-Z}{A}~MeV\nonumber\\
a&=0.5598+0.0014A^{1/3}~fm.
\end{eqnarray}
The database includes 164 even-even nuclei with Z=62-118. The RMS deviation drop to $\sqrt{\frac{\Delta}{164}}=0.190$, which improved $\frac{0.205-0.190}{0.205}=7.3\%$. The calculated results are listed in Table~\ref{Tab3}.

\section{Summary}

In summary, we systematically calculate $\alpha$ decay half-lives for even-even nuclei with proton number from Z=62 to Z=118 within two potential approach based on phenomenological nuclear potential and correction of hindrance factor. A set of new parameters of nuclear potential and analytic expression for hindrance factors is obtained by fitting to the experimental half-lives. Numerical results can well reproduce the experimental half-lives compared with the DDCM and GLDM, eliminating the shortcomings that calculated results deteriorated in the vicinity of the magic number. Finally quantitative results of isospin effect on $\alpha$-core mean nuclear potential have also been given.

\begin{acknowledgments}

This work is supported in part by the National Natural Science Foundation of China (Grant No.11205083), the construct program of the key discipline in hunan province, the Research Foundation of Education Bureau of Hunan Province,China (Grant No.15A159 ),the Natural Science Foundation of Hunan Province,China (Grant No.2015JJ3103),the Innovation Group of Nuclear and Particle Physics in USC, Hunan Provincial Innovation Foundation For Postgraduate (Grant No.CX2015B398).

\end{acknowledgments}

\end{document}